# Superposition States on Different Axes of the Bloch Sphere for Cost-Effective Circuits Realization on IBM Quantum Computers

Ali Al-Bayaty
*Electrical and Computer Engineering Dept.*
*Portland State University*
Oregon, USA
albayaty@pdx.edu

Marek Perkowski
*Electrical and Computer Engineering Dept.*
*Portland State University*
Oregon, USA
h8mp@pdx.edu

*Abstract*—A proposed method for preparing the superposition states of qubits using different axes of the Bloch sphere. This method utilizes the Y-axis of the Bloch sphere using IBM native $\sqrt{X}$ gates, instead of utilizing the X-axis of the Bloch sphere using IBM non-native Hadamard gates, for transpiling cost-effective quantum circuits on IBM quantum computers. In this paper, our presented method ensures that the final transpiled quantum circuits always have a lower quantum cost than that of the transpiled quantum circuits using the Hadamard gates.

## 1 INTRODUCTION

A quantum circuit that intends to solve a classical problem in the quantum domain is the so-called oracle. An oracle is built using a set of quantum gates. These quantum gates are categorized into Boolean gates, e.g., Feynman (CNOT) and $n$-bit Toffoli gates, and Phase gates, e.g., $n$-bit multi-controlled Z (MCZ) gate, where $n \geq 2$. Based on these categorizations of quantum gates that an oracle is constructed from, a quantum circuit is termed either a "Boolean oracle" or a "Phase oracle" [1, 2], depending on the purpose of solving a problem.

Grover's algorithm [1-3], as a quantum search algorithm, is composed of (i) superposition gates, (ii) a Boolean or Phase oracle, and (iii) a Grover diffusion operator. The superposition gates create the complete quantum search space for all $n$ input qubits of an oracle, and the Grover diffusion operator rotates and amplifies the amplitude probabilities of all $n$ input qubits as a true solution (or true solutions) of an oracle, where $n \geq 2$ qubits.

The Hadamard (H) gates are mostly utilized as superposition gates for Grover's algorithm and other search algorithms, such as the quantum random walk search algorithm [4], due to their simplicity of calculation as unitary H matrices [5, 6]. However, for an IBM quantum computer, i.e., a quantum processing unit (QPU), the H gates are IBM non-native gates, which are also termed "non-basis gates". For that, all H gates in a quantum search algorithm need to be decomposed into IBM native gates, and then mapped to an IBM QPU. The native gates of IBM QPU are I, $\sqrt{X}$, X, RZ, and CNOT. In IBM terminologies, the decomposition and mapping processes are termed "transpilation" [7]. Note that if other non-native gates exist in a quantum search algorithm, e.g., $n$-bit Toffoli gates, then these gates need to be transpiled into their equivalent sequences of IBM native gates.

The H gates put a qubit into a superposition state on the X-axis of the Bloch sphere [1, 5, 6]. On the one hand, when a H gate is applied to a qubit of $|0\rangle$ state, then its superposition state becomes $|+\rangle$ or $|0\rangle + |1\rangle$. On the other hand, when a H gate is applied to a qubit of $|1\rangle$ state, then its superposition state becomes $|-\rangle$ or $|0\rangle - |1\rangle$. In contrast, the $\sqrt{X}$ and $\sqrt{X}^\dagger$ gates can be used to put a qubit into a superposition state on the Y-axis of the Bloch sphere [6]. On the one hand, when a $\sqrt{X}$ gate is applied to a qubit of $|0\rangle$ state, then its superposition state becomes $|-i\rangle$ or $|0\rangle - i|1\rangle$. On the other hand, when a $\sqrt{X}$ gate is applied to a qubit of $|1\rangle$ state, then its superposition state becomes $|+i\rangle$ or $|0\rangle + i|1\rangle$. Note that the $\sqrt{X}^\dagger$ gate does the reverse operation of $\sqrt{X}$ gate on the same Y-axis.

For cost-effective transpiling a quantum search algorithm into an IBM QPU, the H superposition gates are replaced by the $\sqrt{X}$ and $\sqrt{X}^\dagger$ gates, since their final transpiled quantum circuits always have a lower quantum cost than that of the transpiled quantum circuits having H gates, i.e., the resulting number of IBM native gates is fewer when using the $\sqrt{X}$ and $\sqrt{X}^\dagger$ superposition gates.

## 2 METHODOLOGY

The Hadamard (H) gate as IBM non-native gate is decomposed into the sequence {RZ(+ π/2) √X RZ(+ π/2)} of IBM native gates. While there is no need to decompose IBM native √X gate. Figure 1(a) illustrates the components of Grover's algorithm [1, 2], and Figure 1(b) illustrates the decomposition of H gates in this algorithm for $n$ input qubits and one Grover loop, where $n \geq 2$. Note that, in the superposition gates, the H superposition gates can also be decomposed into the sequence {RZ(– π/2) √X RZ(+ π/2)} of gates; however, in the Grover diffusion operator, the former and latter H superposition gates should be decomposed into the sequences {RZ(+ π/2) √X RZ(– π/2)} and {RZ(+ π/2) √X RZ(+ π/2)} of gates, to correct the global phase differences.

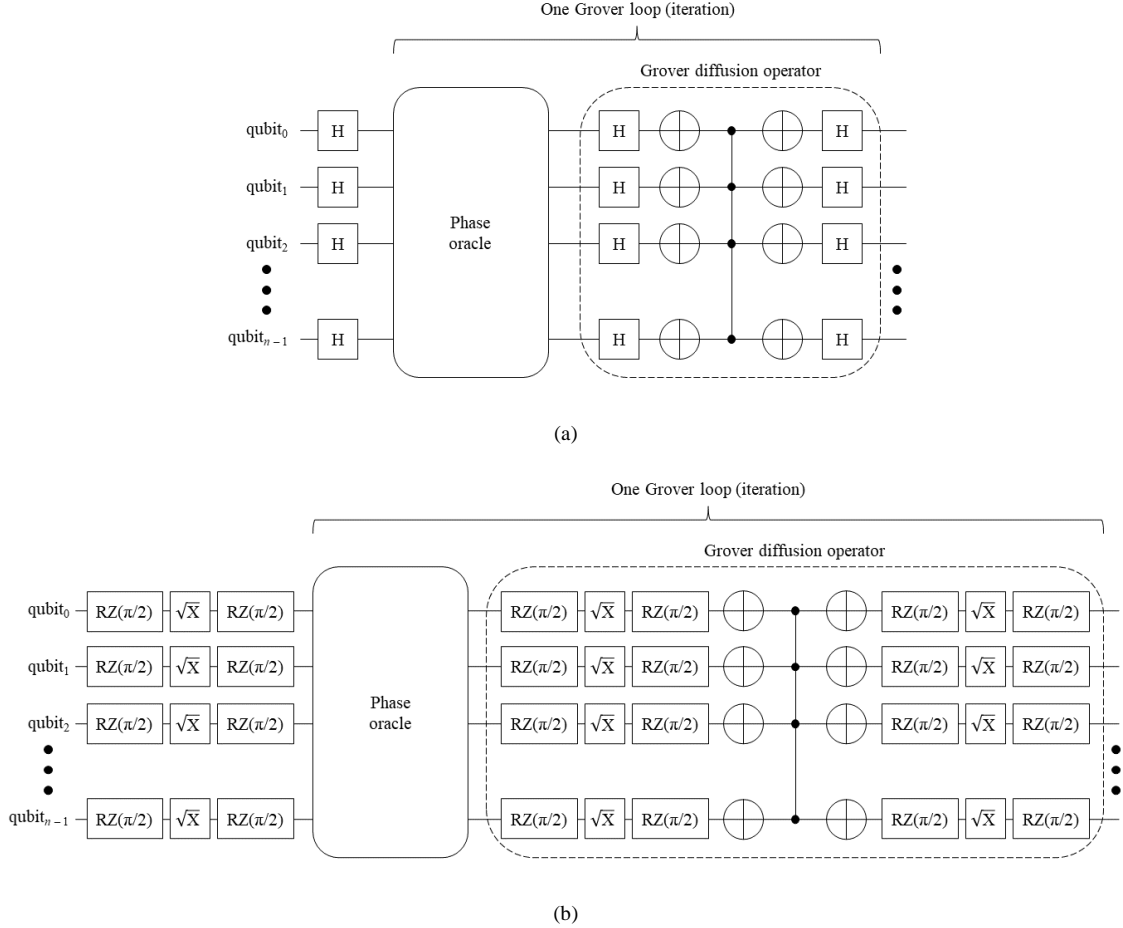

Figure 1. Schematic of the components of Grover's algorithm consisting of superposition gates, a Phase oracle, and Grover diffusion operator using: (a) H superposition gates as IBM non-native gates, and (b) H superposition gates for the sequence {RZ(+ π/2) √X RZ(+ π/2)} of decomposed IBM native gates, for $n$ input qubits and one Grover loop, where $n \geq 2$.

In this paper, the √X and √X$^{\dagger}$ superposition gates replace the H superposition gates for the final lower quantum cost of transpiled quantum circuits on IBM QPUs. Figure 2(a) illustrates such replacements for a Grover's algorithm of $n$ input qubits and one Grover loop, and Figure 2(b) illustrates further reduction of native gates since an X gate is composed of two √X gates, where $n \geq 2$. Note that, in the Grover diffusion operator, the √X$^{\dagger}$ gates replace the former H superposition gates, and the √X gates replace the latter H superposition gates, to correct the final rotational states of a qubit.

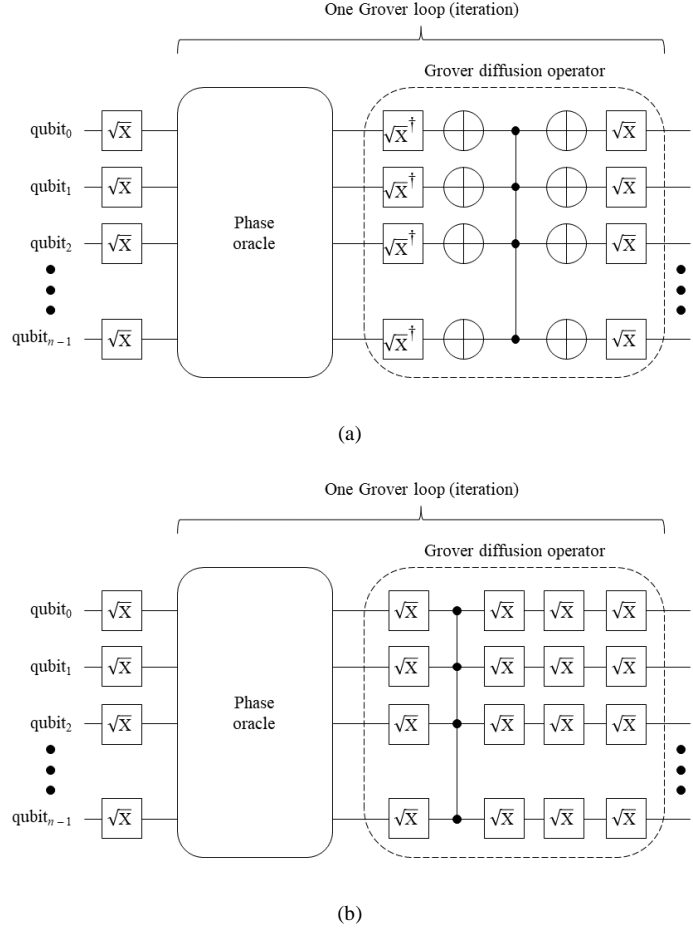

Figure 2. Schematic of the components of Grover's algorithm consisting of superposition gates, a Phase oracle, and Grover diffusion operator using: (a) $\sqrt{X}$ and $\sqrt{X}^\dagger$ gates as replacements for the H superposition gates, and (b) only $\sqrt{X}$ gates as further gates reduction, for $n$ input qubits and one Grover loop, where $n \geq 2$. Note that an X gate is composed of two $\sqrt{X}$ gates, and $\sqrt{X}\sqrt{X}^\dagger = \sqrt{X}^\dagger\sqrt{X} = I$, where I is IBM native identity gate.

As shown in Figure 1(b) and Figure 2(b), for Grover's algorithm, the H superposition gates are reduced from the decomposed $3 \cdot n$ IBM native gates to $n$ IBM native $\sqrt{X}$ gate, and the components of the Grover diffusion operator (except the MCZ gate) are reduced from $8 \cdot n$ IBM native gates to $4 \cdot n$ IBM native $\sqrt{X}$ gates, where $n \geq 2$ input qubits. Such that, the final IBM native gates are reduced from $11 \cdot n$ gates in Figure 1(b) to $5 \cdot n$ gates in Figure 2(b), and for more than 50% gates reduction of a lower transpilation quantum cost [5, 6].

## 3 RESULTS

In our research, Grover's algorithms of 4-bit MCZ oracle are exercised and evaluated, by setting the superposition states of four input qubits on the X-axis of the Bloch sphere (using IBM non-native H gates), as well as setting the superposition states of the same four input qubits on the Y-axis of the Bloch sphere (using IBM native $\sqrt{X}$ gates).

The IBM Quantum Platform was utilized to design and examine these experiments for 1024 shots [8]. However, the technical specifications of T1 (relaxation time), T2 (decoherence time), frequency, anharmonicity, median errors of the native gate, speed of transpilation, number of required pulses, and the speed of Internet connection are not considered for the purpose of our research and experiments.

Figure 3(a) illustrates the quantum circuit of Grover's algorithm for the 4-bit MCZ oracle (Phase oracle) using H superposition gates, and Figure 3(b) demonstrates the outcome probabilities (measurements) of this costly quantum circuit. Figure 4(a) depicts the quantum circuit of Grover's algorithm for the 4-bit MCZ oracle (Phase oracle) using $\sqrt{X}$ superposition gates, and Figure 4(b) shows the outcome probabilities of this cost-effective quantum circuit. From Figure 3(b) and Figure 4(b), both outcome probabilities are identical, which means that IBM native $\sqrt{X}$ superposition gates are successfully operated by replacing costly IBM non-native H gates. Note that, in the Phase oracle and Grover diffusion operator, the 4-bit MCZ gate is composed of a 4-bit Toffoli gate that is surrounded by two H gates for phases inversion and not for the purpose of superposition; for this reason, these two H non-native gates are not replaced by the $\sqrt{X}$ native gates.

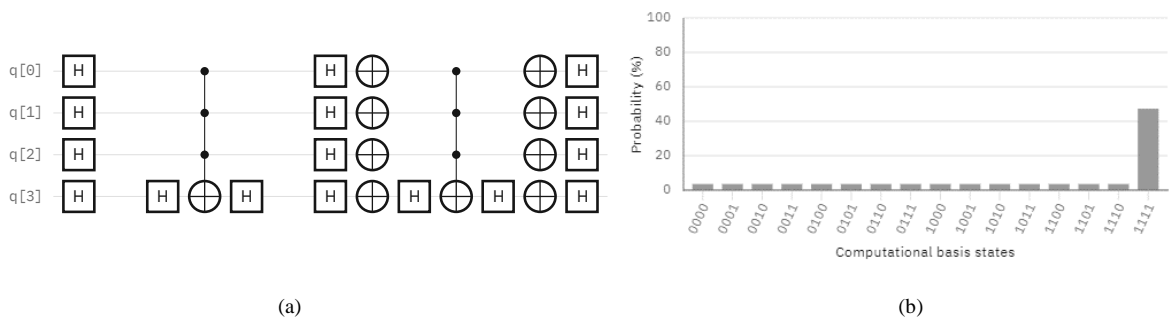

(a) (b)

Figure 3. Grover's algorithm of 4-bit MCZ oracle (Phase oracle): (a) the complete quantum circuit using IBM non-native H superposition gates, and (b) the outcome probabilities (measurements) of this costly quantum circuit.

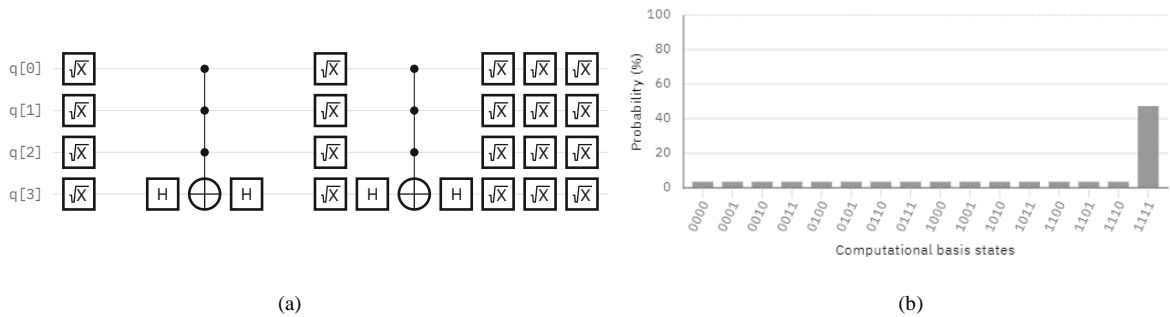

(a) (b)

Figure 4. Grover's algorithm of 4-bit MCZ oracle (Phase oracle): (a) the complete quantum circuit using IBM native $\sqrt{X}$ superposition gates, and (b) the outcome probabilities (measurements) of this cost-effective quantum circuit.